\documentclass[twocolumn,floatfix,aps,pra,showpacs]{revtex4-1}
\usepackage{multirow}
\usepackage{graphicx}
\usepackage{tabularx}
\usepackage{booktabs}
\usepackage{epsfig,graphicx}%[draft]
\usepackage{flafter}
\usepackage{amsmath}
\usepackage{color}
\usepackage{braket}

\usepackage[colorlinks,urlcolor=blue,citecolor=blue,linkcolor=blue]{hyperref}

\begin{document}

\title{Higher-nodal collective modes in a resonantly interacting Fermi gas}

\author{Edmundo R. S\'anchez Guajardo}\altaffiliation[Present address: ]
{Institute of Molecular Pathology, Vienna, Austria.}
\author{Meng Khoon Tey}\altaffiliation[Present address: ]
{Tsinghua University, Beijing, China.}
\author{Leonid A. Sidorenkov}
\author{Rudolf Grimm}
\affiliation{Institut f\"ur Quantenoptik und Quanteninformation (IQOQI),
 \"Osterreichische Akademie der Wissenschaften}
 \affiliation{Institut f\"ur Experimentalphysik, Universit\"at Innsbruck, 6020 Innsbruck, Austria}

\date{\today}

\begin{abstract}
We report on experimental investigations of longitudinal collective oscillations in a highly elongated, harmonically trapped two-component Fermi gas with resonantly tuned $s$-wave interactions (`unitary Fermi gas'). We focus on higher-nodal axial modes, which in contrast to the elementary modes have received little attention so far. We show how these modes can be efficiently excited using a resonant local excitation scheme and sensitively analyzed by a Fourier transformation of the detected time evolution of the axial density profile. We study the temperature dependence of the mode frequencies across the superfluid phase transition. The behavior is qualitatively different from the elementary modes, where the mode frequencies are independent of the temperature as long as the gas stays in the hydrodynamic regime. Our results are compared to theoretical predictions based on Landau's two-fluid theory and available experimental knowledge of the equation of state. The comparison shows excellent agreement and thus both represents a sensitive test for the validity of the theoretical approach and provides an independent test of the equation of state. The present results obtained on modes of first-sound character represent benchmarks for the observation of second-sound propagation and corresponding oscillation modes.
\end{abstract}
\date{\today}

\maketitle

\section{Introduction}
In ultracold quantum gases, measurements on collective oscillations are well established as powerful tools to study the many-body properties of the system \cite{Pitaevskii2003book, Pethick2002book}. Experiments on collective modes reveal the dynamics in the different regimes of superfluid, collisionally hydrodynamic, and collisionless behavior. The eigenfrequencies can be determined very accurately, which allows to extract valuable information on the equation of state (EOS), with great sensitivity to subtle interaction effects in the strongly interacting regime.

In ultracold Fermi gases \cite{Varenna2008ufg, Giorgini2008tou, Bloch2008mbp}, collective modes have been widely applied to study the crossover from Bose-Einstein condensation (BEC) to a Bardeen-Cooper-Shrieffer (BCS) type superfluid. A situation of particular interest is the two-component Fermi gas with resonant interactions, with an $s$-wave scattering length tuned to infinity by means of a Feshbach resonance \cite{Chin2010fri}. This special case, which lies right in the center of the BEC-BCS crossover, has attracted a great deal of interest, mainly attributed to its universal properties. The resonantly interacting Fermi gas is characterized by strong interaction effects in the EOS \cite{Kinast2005hco, Horikoshi2010mou, Nascimbene2010ett, Ku2012rts} and reveals a unique universal thermodynamic behavior \cite{Ho2004uto}.

So far, experiments on collective modes in harmonically trapped Fermi gases have been restricted to a few elementary modes. The most simple modes, sloshing modes, do not provide any information on the properties of the quantum gas, and their main application is thus to accurately determine the trap frequencies. Surface modes are insensitive to the EOS, but they allow to clearly distinguish between hydrodynamic and collisionless behavior \cite{Altmeyer2007doa, Wright2007ftc} and they have been used to detect the angular momentum in a rotating Fermi gas \cite{Riedl2009loa, Riedl2011sqo}. Elementary compression modes of axial \cite{Bartenstein2004ceo, Nascimbene2009coo} and radial \cite{Bartenstein2004ceo, Kinast2004efs, Kinast2004boh, Kinast2005doa, Altmeyer2007pmo, Riedl2008coo} character have been very widely studied in the field. Such modes do not only probe the particular collision regime, but they also give access to the compressibility of the gas. However, for a unitary Fermi gas, the eigenfrequencies of the simple compression modes do not show any variation across the superfluid phase transition as the temperature is varied \cite{Kinast2005doa, Riedl2008coo}. This can be understood as a consequence of the fact that superfluid and collisional hydrodynamics both lead to the same frequencies. A rigorous proof for this temperature-independence can be given in terms of an exact scaling solution of the hydrodynamic equations of motion \cite{Hou2013sso}. The situation is strikingly different for higher-nodal modes. Here the frequencies vary across the superfluid phase transition, when the dynamical regime changes from superfluid to collisional hydrodynamics \cite{Tey2013cmi, Hou2013fas}. Such higher-nodal modes therefore represent an interesting addition to the experimental tool-box for probing strongly interacting Fermi gases.

We have recently carried out a series of experiments on higher-nodal axial modes in the geometry of a highly elongated trapping potential. First results on the temperature dependence have already been presented in Ref.~\cite{Tey2013cmi}, and the general theoretical framework is described in Ref.~\cite{Hou2013fas}. In this Article, we briefly summarize the main theoretical predictions (Sec.~\ref{sec:theory}), we describe the experimental procedures in more detail (Sec.~\ref{sec:procs}), and we present the whole set of our experimental results obtained for two different higher-nodal modes (Sec.~\ref{sec:results}). While we here restrict our attention to modes of first-sound character, we note that the results are important as benchmarks for the observation of second-sound propagation \cite{Sidorenkov2013ssa} and in view of future experiments on second-sound modes (Sec.~\ref{sec:outlook}).

%Below the critical temperature $T_c$, one expects two types of collective modes corresponding to standing waves of first and second sound in a superfluid~\cite{Khalatnikov1965ait}. The first-sound collective modes manifest themselves as density oscillations and they also exist above $T_c$. In comparison, the second-sound modes essentially correspond to entropy oscillations, which are much more difficult to detect. Here we restrict our attention to first-sound modes.

\section{Theoretical predictions}\label{sec:theory}
%Our focus here is the higher-nodal collective oscillations exhibited by a resonantly-interacting Fermi gas at finite temperature.

Higher-nodal collective modes in Fermi gases have been theoretically studied based on Landau's two-fluid equations  for an isotropic harmonic trapping geometry \cite{He2007fas, Taylor2009fas}. For real experiments, however, the situation of highly elongated harmonic traps is more relevant. In this geometry, the description can be reduced to a set of effectively 1D hydrodynamic equations, which only depend on the axial coordinate $z$. This simplification leads to a powerful approach to describe sound propagation and collective modes in experimentally realistic situation. The basic approach was introduced in Ref.~\cite{Bertaina2010fas} for a cylindrical trap geometry with tight radial confinement. Reference~\cite{Hou2013fas} presents a generalization to the situation of additional weak axial confinement, which readily describes the commonly used geometry of a highly elongated trap containing a `cigar-shaped' atomic sample. Here, we summarize the main elements of this theoretical approach and the corresponding predictions for higher-nodal modes of first-sound character.

The two basic assumptions underlying the 1D hydrodynamic approach are thermal equilibrium in the radial direction and a flow field that is independent of the radial position. This corresponds to conditions of sufficient heat conductivity and sufficient shear viscosity, which are readily satisfied for resonantly interacting Fermi gases under common trapping conditions. Applying the local density approximation, one can describe the thermodynamics and the flow properties of the trapped sample using effective 1D quantities, which are derived by integrating over the transverse degrees of freedom, such that a thermodynamic quantity $q$ yields a 1D counterpart $q_1 = 2\pi\int_0^{\infty}q\,r\,dr$.

For a first-sound collective mode with frequency $\omega$, the local flow speed can be expressed as $v(z,t) = v_z(z) e^{-i\omega t}$, where the $z$-dependent amplitude $v_z(z)$ represents the spatial oscillation of the flow velocity. The hydrodynamic equation that describes $v_z(z)$ takes the form~\cite{Hou2013fas}
\begin{equation}\label{HDT}
m(\omega^{2}-\omega^2_z)v_z-\frac{7}{5}m\omega^2_zz\partial_zv_z +\frac{7}{5}\frac{P_1}{n_1}\partial_z^2v_z =0\,.
\end{equation}
Here, $\omega_z$ represents the trap frequency along the axial direction, $m$ is the atomic mass, $P_1$ is the `1D pressure' (having units of force), and $n_1$ is the linear number density. The equation is valid for small-amplitude oscillations, which can be treated as perturbations by linearizing Landau's equations.

At zero temperature and in the classical limit of high temperature the hydrodynamic equation (\ref{HDT}) admits analytic solutions of polynomial form $v_z = a_kz^k +a_{k-2}z^{k-2}+...$, with integer values of $k$. At $T=0$,  where $P_1(n_1)/n_1 =(2/7)[\mu_0 -(1/2)m\omega^2_zz^2]$, with $\mu_0$ being the chemical potential at the center of the trap, the frequency of the $k$-th mode is given by
\begin{equation}
\omega^2=\frac{1}{5}(k+1)(k+5)\omega_z^2 \, .
\label{HD3}
\end{equation}
In the classical limit, where $P_1/n_1=k_{B}T$, one finds the different $k$-dependence
\begin{equation}
\omega^2=\frac{1}{5}(7k+5)\omega_z^2 \, .
\label{HD4}
\end{equation}

We point out that Eqs.~(\ref{HD3}) and (\ref{HD4}) give the same values for $k=0$ (center of mass oscillation) and $k=1$ (lowest axial breathing mode). In fact, one can prove that the frequencies of these two lowest modes are temperature independent for the resonantly interacting gas (unitary Fermi gas), corresponding to an exact scaling solution of the two-fluid hydrodynamic equation~\cite{Hou2013sso}. On contrary, the frequencies of the $k\ge 2$ modes vary with temperature.

We now focus on the modes with $k=2$ and $k=3$, which are experimentally most relevant. Using a variational approach~\cite{Taylor2008vto}, one can obtain their eigenfrequencies at finite temperatures as
\begin{equation}
\omega^2_{k=2}= \frac{129t_2-25}{5(9t_2-5)}\omega^{2}_{z}\,,
\label{eqn:fk2}
\end{equation}
and
\begin{equation}
\omega^2_{k=3}=\frac{440 t_3 -252}{5(25t_3-21)}\omega^{2}_{z}\,.
\label{eqn:fk3}
\end{equation}
In these equations $t_2= M_0M_4/M_2^2$ and $t_3= M_2M_6/M_4^2$, where we have introduced the dimensionless moments
 \begin{equation}
M_\ell= \int_{-\infty}^{\beta\mu_0}dx(\beta\mu_0-x)^{(\ell+1)/2}f_n(x).
\label{Ml}
\end{equation}
Here, the phase-space density $f_n(x)$ is a universal function \cite{Ho2004uto} defined by $f_n(x) = n\lambda_T^3$, where $n$ is the 3D number density and $\lambda_T = h/(2\pi m k_B T)^{1/2}$ is the thermal deBroglie wavelength.  The dimensionless parameter $x = \beta\mu$, with $\beta = 1 /k_B T$ and $\mu$ being the chemical potential, is related uniquely to $T/T_F$. The universal function $f_n(x)$ can be determined from the recent EOS measurements~\cite{Kinast2005hco, Horikoshi2010mou, Nascimbene2010ett, Ku2012rts}. In this work, we make use of the latest results from~\cite{Ku2012rts}.

One can also show that the velocity fields for the $k=2$ and $k=3$ modes take the form
\begin{equation}\label{eqn:k2velocity}
v_z^{k=2}(z) \propto \frac{3m\omega_z^2\beta}{2}\frac{M_0(x_0)}{M_2(x_0)}z^2 - 1\,,
\end{equation}
and
\begin{equation}\label{eqn:velocity}
v_z^{k=3}(z) \propto \frac{5m\omega_z^2\beta}{6}\frac{M_2(x_0)}{M_4(x_0)}z^3 - z\,.
\end{equation}
Here, the parameter $x_0=\beta\mu_0$ is the value of $x$ at the center of the trap. Finally, using the equation of continuity under the 1D formulation $\partial_t n_1+ \partial_z(n_1 v_z)=0$, one can calculate the shape of the density oscillations for each mode.

\section{Experimental procedures}
\label{sec:procs}

\subsection{Sample preparation}

The starting point of our experiment is an ultracold, resonantly interacting Fermi gas in an elongated optical dipole trap. This gas is prepared by evaporating a balanced mixture of fermionic $^6$Li atoms in their two lowest spin states at a magnetic field of 834\,G, very close to the center of the well-known broad Feshbach resonance~\cite{Chin2010fri, 832note}. The atomic cloud contains typically $N/2=1.5\times 10^5$ atoms per spin state. For the lowest temperatures, the waist of the trapping beam (wavelength 1075 nm) is 31~$\mu$m, the trap depth is about $2\,\mu$K, and the axial and radial trap frequencies are $\omega_z=2\pi\times 22.52(2)\,$Hz and $\omega_r=2\pi\times 473(2)\,$Hz, respectively. For experiments at higher temperatures, the beam waist is increased to 38~$\mu$m, and deeper traps are used (up to $16\,\mu$K depth) with trap frequencies of up to $\omega_z=2\pi\times 23.31(3)\,$Hz and $\omega_r=2\pi\times 1226(6)\,$Hz. The corresponding Fermi temperatures $T_F=\hbar(3N\omega_r^2\omega_z)^{1/3}/k_B$ vary between about $0.8$ and $1.5\,\mu$K.

To achieve the lowest possible temperatures, we perform deep evaporative cooling up to the point where the trapping potential cuts slightly into the Fermi sea, indicated by the onset of spilling losses in the last stage of the evaporation. After that, the gas is adiabatically recompressed by increasing the trapping beam's power to the extent where the trap depth becomes at least twice more than the Fermi energy $k_B T_F$. This recompression step is essential to ensure negligible anharmonicities in the radial confinement. The essentially perfect harmonic confinement along the axial direction is ensured by the magnetic trapping that results from the curvature of the magnetic field used for Feshbach tuning~\cite{Jochim2003bec}.

We vary the temperature $T$ of the gas by controlled heating, always starting from a deeply cooled cloud ($T/T_F \approx 0.1$). In the low-temperature range ($T \lesssim 0.2\,T_F$), we simply introduce a variable hold time of up to 4\,s in which residual trap heating slowly increases the cloud's temperature. For the higher-temperature range ($0.2\,T_F < T < 0.5\,T_F$), we heat the sample using parametric heating, modulating the trap power at about $2 \omega_r$, and introducing a sufficient hold time to reach thermal equilibrium between the different degrees of freedom. We note that we use deeper traps for samples with higher $T/T_F$, because plain evaporation puts a limit on the maximum attainable temperature of the gas.

\subsection{Thermometry}

We determine the temperature $T$ of the gas by analyzing its density distribution in the trap, based on knowledge of the EOS. Under the local density approximation, one can readily show for a harmonic trap that the 1D density profile $n_1(z)$ is given by~\cite{Nascimbene2010ett, Ho2010otp}
\begin{equation}\label{eqn:n1}
n_1(z) = \frac{2\pi}{m\omega_r^2}\frac{k_B T}{\lambda_T^3}f_p(x_0 - \frac{1}{2}\beta m\omega_z^2z^2) \, .
\end{equation}
Here, the universal function $f_p(x)$ is related to the universal function $f_n(x)$ introduced in Sec.~\ref{sec:theory} by $f_p(x) = \int_{-\infty}^x f_n(y)\,dy$, and is therefore also known from a given EOS. The parameter $x_0$ is the value of $x$ at the center of the trap.

In the ideal case where one is able to obtain an accurate {\it in-situ} measurement of $n_1(z)$ by imaging the trapped cloud, it is straightforward to retrieve the parameters $T$ and $x_0$ by fitting $n_1(z)$ using Eq.~(\ref{eqn:n1}). However, in reality we have to deal with imperfections of our absorption imaging scheme. To extract the temperature in an accurate way, we have adopted the methods described in detail in the Appendix.

\subsection{Exciting and observing higher-nodal collective modes}

\begin{figure}
\includegraphics[width=0.8 \columnwidth]{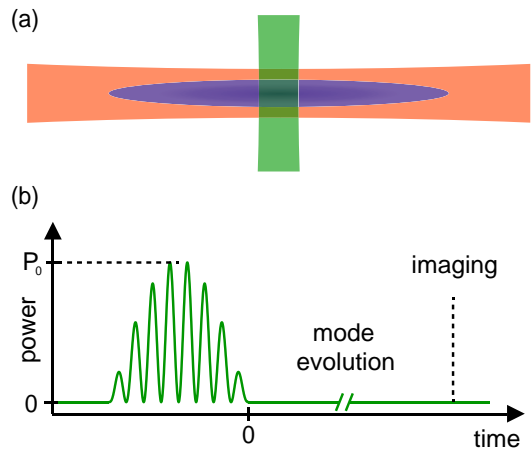}
\caption{\label{figure:excitation_scheme} (Color online) Experimental scheme to excite higher-nodal first-sound longitudinal modes. In (a), we illustrate the basic geometry of exciting the optically trapped cloud with a weak, power-modulated repulsive laser beam, which perpendicularly intersects the trapping beam. In (b), we show the power modulation of the repulsive beam for the excitation.}
\end{figure}

We apply a resonant excitation scheme to create a collective oscillation. Figure~\ref{figure:excitation_scheme}(a) illustrates the basic geometry of our scheme in which a repulsive 532-nm green laser beam perpendicularly intercepts the trapping beam. To excite a mode, we position the green beam near the antinode of the mode and modulate its power at the expected frequency of the mode. The amplitude, duration and shape of the modulation are carefully adjusted in order not to overdrive the excitation while maintaining sufficient signal-to-noise ratio. %Overdriving the mode can cause heating and non-linear response in the oscillation dynamics, which would shift the mode frequency.
%After optimization,
We adopt an excitation pulse that contains 8 cycles of sinusoidal modulation with a half-cycle sine envelope \cite{pulseComment}, see illustration in Fig.~\ref{figure:excitation_scheme}(b). We set the maximum barrier height of the green beam to about $0.01\,k_B T_F$. Depending on the order of the mode to be excited, this is realized with beam waists ranging from $30\,\mu$m to $70\,\mu$m and values of the maximum power $P_0$ between $400\,\mu$W and 3\,mW.
%with a typical power of $P_0 = 0.5$\,mW.
The 8-cycle pulse is chosen such that the resulting total excitation duration is not too long as compared to the damping time of the highest-nodal mode that we can observe. The smooth half-cycle sine envelope reduces the Fourier width and avoids side lobes in the spectrum, thereby suppressing the excitation of unwanted modes. For an efficient excitation of a given mode, we find that the width of the green beam should well match the local mode profile at the selected antinode, while the excitation frequency should be within 1\% of the actual mode frequency.

\begin{figure*}
\includegraphics[width=0.75\textwidth]{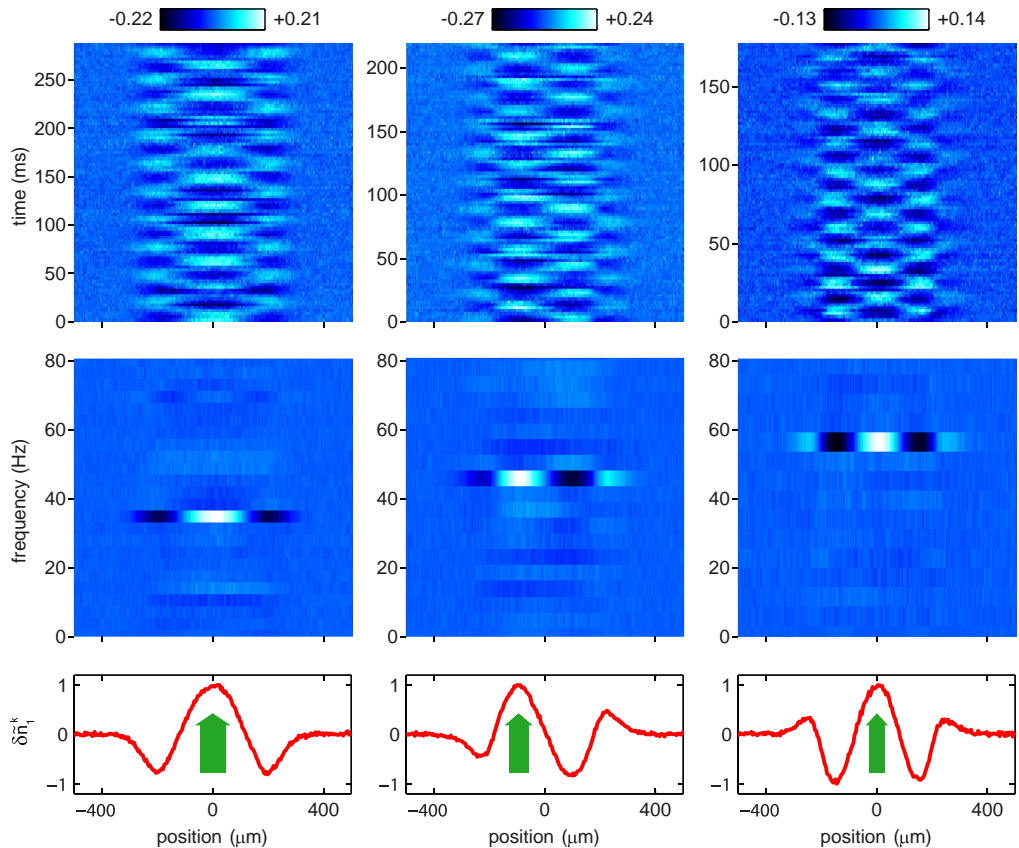}
\caption{\label{figure:3modes} (Color online) Observed collective modes in the time-domain and in frequency space. The three columns refer to the mode orders $k=1, 2$, and 3. We show the normalized density variation signals $\delta n_1(z, t)/\bar{n}_1(0)$ (top row), their Fourier transforms $\delta \tilde{n}_1(z,\omega)$ (middle row), and the mode profiles $\delta \tilde {n}_1^k$ (bottom row). The measurements were taken for our coldest samples with $T/T_F\approx 0.1$. The arrows in the bottom panel show the positions of the repulsive excitation beam for each mode. Note that for efficient excitation we adjust the width of the excitation beam to match the local profile at the chosen antinode. While, for $k=1$ and $3$, the beam is centered and addresses the central antinode, the beam is spatially offset for the $k=2$ mode.}
\end{figure*}

Once a collective mode is excited, we record the axial density profiles $n_1(z,t)$ of the gas for a variable time delay $t$ after the excitation pulse, where $n_1(z,t)$ is the number density integrated over the transverse degrees of freedom. These profiles are obtained with a probe beam that perpendicularly intercepts the trapping beam, and are taken 600\,$\mu$s after suddenly releasing the atoms from the optical trap~\cite{TOFComment}. To enhance the visibility, we subtract a background profile $\bar{n}_1(z)$ obtained from averaging the profiles over all measured delay times. This gives a differential density variation function $\delta n_1(z, t) = n_1(z,t) - \bar{n}_1(z)$, which is finally normalized to the maximum linear density $\bar{n}_1(0)$ at the trap center. In the top panel of Fig.~\ref{figure:3modes}, we show examples of this signal for the $k=1$, 2, and 3 modes for the coldest samples with $T/T_F\approx0.1$. One can see that the adjacent antinodes always oscillate in opposite directions, similar to standing waves on a guitar string.

\subsection{Analyzing the eigenmodes: Extracting mode profiles, frequencies, and damping rates}

The first step to analyze the observed time-dependent profiles $\delta n_1(z,t)$ is a Fourier transform, yielding a representation of our data in frequency space, $\delta \tilde{n}_1(z,\omega)$~\cite{FFTComment}. For this purpose we  employ a fast Fourier transform (FFT) algorithm.
%To reveal the eigenfrequencies and profiles of collective modes being excited, we perform fast Fourier transform (FFT) on $\delta n_1(z,t)$ to obtain.
Corresponding results are shown in the middle panel of Fig.~\ref{figure:3modes}, as calculated for each time-dependent oscillation profile in the top panel. The discrete nature in frequency space becomes evident, with very little noise in the background.

It is straightforward to extract the mode profiles $\delta \tilde{n}_1^k(z)$ from the FFT results by setting $\delta \tilde{n}_1^k(z)=\delta \tilde{n}_1(z,\omega_k)$, where $\omega_k$ is the eigenfrequency of the $k^\mathrm{th}$ mode. The corresponding mode profiles for the $k=1$, 2, and 3 modes are shown in the bottom panel of Fig.~\ref{figure:3modes}. Experimentally, we make use of the mode profiles and the frequencies obtained in this way to optimize the beam waist and the modulation frequency in our excitation scheme. We proceed iteratively, which eventually allows for an optimum excitation of a single mode.

To extract the mode frequencies more precisely than it is possible by simply analyzing the peaks in the corresponding Fourier spectrum, we adopt the following algorithm. We project $\delta n_1(z, t)$ onto the mode profile $\delta \tilde{n}_1^k(z)$ to obtain a mode amplitude function $A(t)=\int_{-\infty}^{\infty} \delta n_1(z,t) \delta \tilde{n}_1^k(z)\,dz$. Then we fit a simple damped harmonic oscillation to $A(t)$ to obtain the frequency and the damping time of the mode. This projection procedure is analogous to the projection of a superposition wavefunction onto one of the orthogonal eigenstates of a quantum system. It greatly enhances the signal-to-noise ratio and results in very low statistical uncertainties for the mode frequencies, with relative uncertainties as low as in the range of a few permille.

\begin{figure}
\includegraphics[width=0.8\columnwidth]{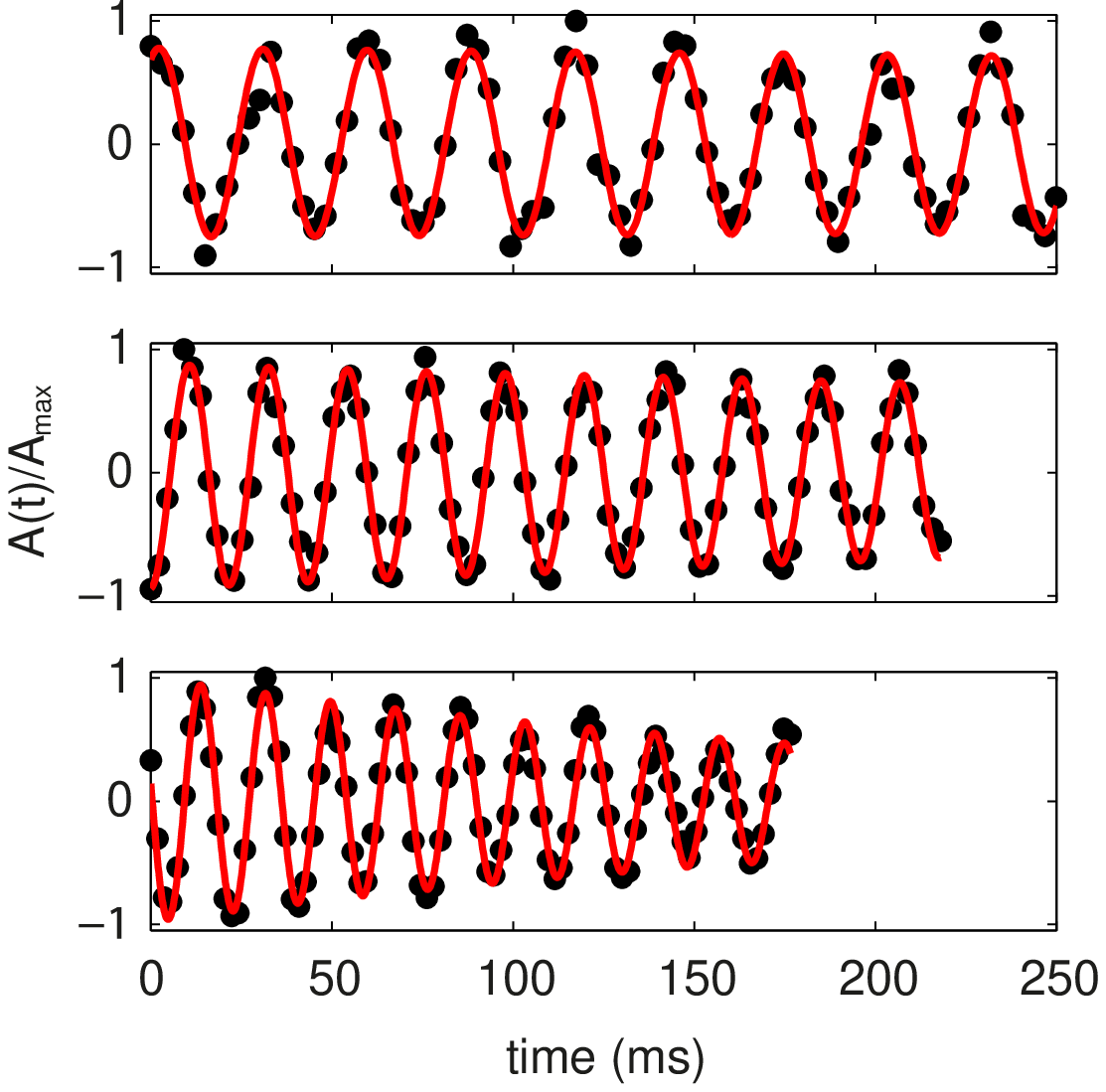}
\caption{\label{figure:damping} Evolution of the $k=1$ (top), $k=2$ (middle) and $k=3$ (bottom) mode amplitudes at $T/T_F\approx 0.1$. The black circles represent the experimental data for $A(t)$ normalized to the maximum value $A_\mathrm{max}$ of $|A(t)|$. The red solid lines are the fits to the data based on simple damped harmonic oscillations.}
\end{figure}

In Fig.~\ref{figure:damping}, we show examples of $A(t)$ for the $k=1$, 2, and 3 modes obtained from samples with $T/T_F\approx0.1$. For the $k=1$ compression mode (upper panel), the observed behavior does not show any damping. Even data taken after a much longer delay time of 3\,s (not shown) do not reveal any significant damping. In contrast, the higher-nodal modes show clear damping. At the lowest temperatures, the 1/$e$ damping time for the $k=2$ mode (middle panel) is about 1.5\,s, and that for the $k=3$ mode (lower panel) is about 0.3\,s.

We finally note that we have not succeeded in observing modes with $k \ge 4$, despite of considerable efforts. We believe that this is due to a fast increase of damping with the mode order, which is clearly indicated by our data for the $k=1, 2, 3$ modes. Large damping affects both our resonant excitation scheme and the detection of the mode by means of a Fourier transform, which may explain a huge difference between the observed mode with $k=3$ and the unobserved mode with $k=4$.

\subsection{Checking for systematic errors}

The real experiment is an approximation to the ideal scenario of a small-amplitude oscillation in a perfectly harmonic trap, as described in the theoretical approach in Sec.~\ref{sec:theory}. Here we investigate in how far our data are influenced by anharmonicities of the trapping potential and nonlinear effects arising from the finite amplitude of the mode, and we identify the conditions that ensure a reliable comparison between the measurements and the theoretical predictions.

The axial compression mode ($k=1$) serves us as a benchmark to rule out a significant effect of anharmonicities. This mode has been studied extensively before~\cite{Bartenstein2004ceo, Nascimbene2009coo} and, in the unitarity limit, its frequency is temperature independent as long as the gas remains hydrodynamic \cite{Hou2013sso}. Only for very shallow traps, we observe a $k=1$ mode frequency that is lower than expected. We find that trap depths of twice the non-interacting Fermi energy are sufficient to observe a frequency very close to the ideal value of $\omega_{k=1} = \sqrt{12/5} \, \omega_z$ throughout the full temperature range explored in the present work. Deviations from this value remain below 0.3\% and no significant temperature dependence is observed.

%The theoretical results presented in Sec.~\ref{sec:theory} are derived with the assumption of small perturbations, which are necessary for linearizing the Landau's two-fluid equations.

\begin{figure}
\includegraphics[width=0.8\columnwidth]{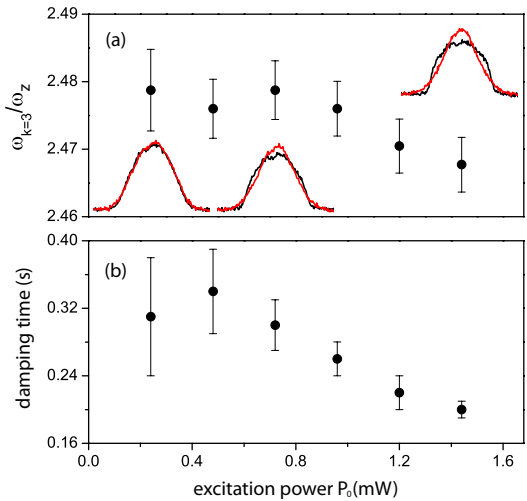}
\caption{\label{figure:overdrive} The normalized oscillation frequency (a) and 1/$e$ damping time (b) of the $k=3$ mode versus the power $P_0$ of the green excitation beam \cite{pulseComment}. The three insets in (a) show the axial density profiles of the atom cloud at the turning points of the mode oscillation for three different excitation powers. The error bars denote the standard errors obtained from fitting $A(t)$.}
\end{figure}

We checked for a possible nonlinear behavior by deliberately overdriving the collective modes. We measured the frequency and damping time of each collective mode versus the power of the excitation beam. Figure~\ref{figure:overdrive} shows the results of such a measurement for the $k=3$ mode for the coldest samples used in our work. The measurements show that the frequency stays constant within the measurement uncertainties up to a power $P_0 \approx 1$\,mW. The fact that the spatial profile is already strongly affected (see insets) shows that the mode frequency is rather robust against nonlinearities. The mode damping time exhibits a similarly robust behavior with a slight trend to be affected already at somewhat smaller excitation amplitudes. The excitation power $P_0$ is always kept small enough to avoid significant effects on the mode frequencies.

\section{Experimental results}
\label{sec:results}

\begin{figure}
\includegraphics[width=0.95\columnwidth]{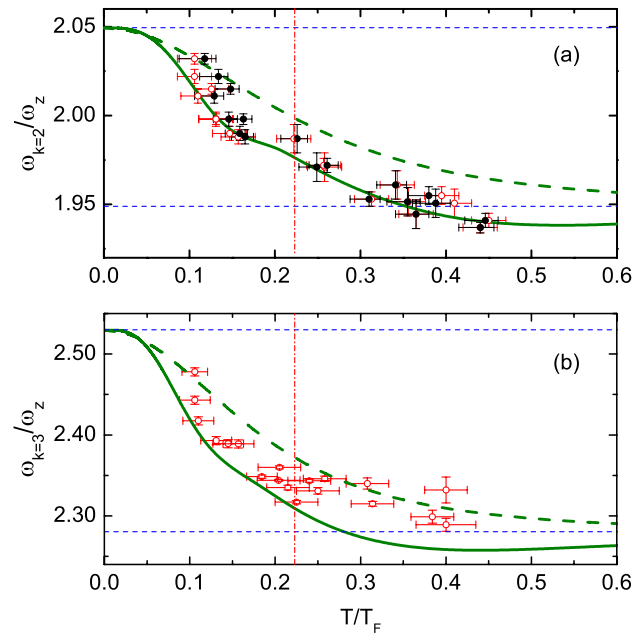}
\caption{\label{fig:k2k3freq_TTF} (Color online) Comparison between experimental and theoretical first-sound frequencies for (a) the $k=2$ and (b) the $k=3$ mode. In (a), the reduced temperature $T/T_F$ is obtained by the two different methods described in the Appendix in Secs.~\ref{subsec:TTF} (open red symbols) and \ref{subsec:EE0} (filled black symbols). In (b), $T/T_F$ is extracted by only using the first method. The theoretical curves (solid lines) are obtained with Eqs.~(\ref{eqn:fk2}) and (\ref{eqn:fk3}) using the EOS of Ref.~\cite{Ku2012rts}. For comparison, we also show the mode frequencies (dashed curves) that would result from the same equations but using the EOS of the ideal Fermi gas. In both panels, the upper and lower thin horizontal dashed lines mark the zero-$T$ superfluid limits and the classical hydrodynamic limits whose values are given by Eqs.~(\ref{HD3}) and (\ref{HD4}), respectively. The dash-dot vertical lines in (a) and (b) indicate the critical temperature $T_c/T_F = 0.223(15)$.}
\end{figure}

Here we present our main experimental results and compare them with the predictions of Sec.~\ref{sec:theory}. We consider the two modes with $k=2$ and $k=3$ and discuss how their eigenfrequencies and the corresponding mode profiles depend on the temperature. We furthermore present data on the temperature-dependent damping of the two modes.

In Fig.~\ref{fig:k2k3freq_TTF}(a) and (b), we show the measured mode frequencies $\omega_{k=2}$ and $\omega_{k=3}$ versus temperature, normalized to the axial trap frequency $\omega_z$. The two limiting cases of a $T=0$ superfluid and a classical collisionally hydrodynamic gas are indicated by the upper and the lower horizontal dashed lines in both panels, see Eqs.~(\ref{HD3}) and (\ref{HD4}). The theoretical predictions according to Eqs.~(\ref{eqn:fk2}) and (\ref{eqn:fk3}) and the EOS from Ref.~\cite{Ku2012rts} are shown by the solid lines. For comparison, the hypothetical frequencies calculated with the EOS of a non-interacting Fermi gas are shown by the dashed lines. For the $k=2$ mode we have applied both thermometry methods as described in the Appendix, with the open symbols representing the results from the wing-fit method (Appendix Sec.~\ref{subsec:TTF}) and the filled symbols from the potential-energy method (Appendix Sec.~\ref{subsec:EE0}). For the $k=3$ mode we have applied only the first method.

For the $k=2$ mode (Fig.~\ref{fig:k2k3freq_TTF}(a)), the measured mode frequencies are in almost perfect agreement with the theoretical predictions based on the EOS from~\cite{Ku2012rts}. In comparison, the disagreement with the dependence that would result from the EOS of the ideal Fermi gas highlights the important role of interactions. At the lowest temperature realized in our experiment ($T/T_F \approx 0.1$), the frequency lies close to the $T=0$ superfluid limit ($\omega_{k=2}/\omega_z = 2.049$), but already shows a significant down-shift amounting to almost 1\%. This illustrates the high sensitivity of the mode frequency to finite-temperature effects. At the highest temperatures ($T/T_F \approx 0.45$) our data show a clear trend to go below the asymptotic high-temperature value ($\omega_{k=2}/\omega_z = 1.949$), which corresponds to the classical hydrodynamic case. This nonmonotonic temperature dependence can be understood based on the first-order correction to the EOS resulting from the virial expansion \cite{Liu2009vef} at high temperatures.

For the $k=3$ mode (Fig.~\ref{fig:k2k3freq_TTF}(b)), the general behavior is very similar to the $k=2$ mode, with the main difference that the relative frequency change from superfluid to collisional hydrodynamics ($\omega_{k=2}/\omega_z = 2.530$ and $2.280$, respectively) is about two times larger. For temperatures below $0.2\,T_F$ we find similarly good agreement as in the $k=2$ case. However, for higher temperatures there is a significant trend to lie above the predicted frequencies, so that we never observe values below the high-temperature limit. This discrepancy is most likely due to dissipative effects, which manifest themselves in mode damping. Landau's equations used as a basis for describing the collective modes do not contain dissipative terms and can therefore not make any predictions on the damping behavior. We can therefore address the question of damping only from the experimental side.

\begin{figure}
\includegraphics[width=0.8\columnwidth]{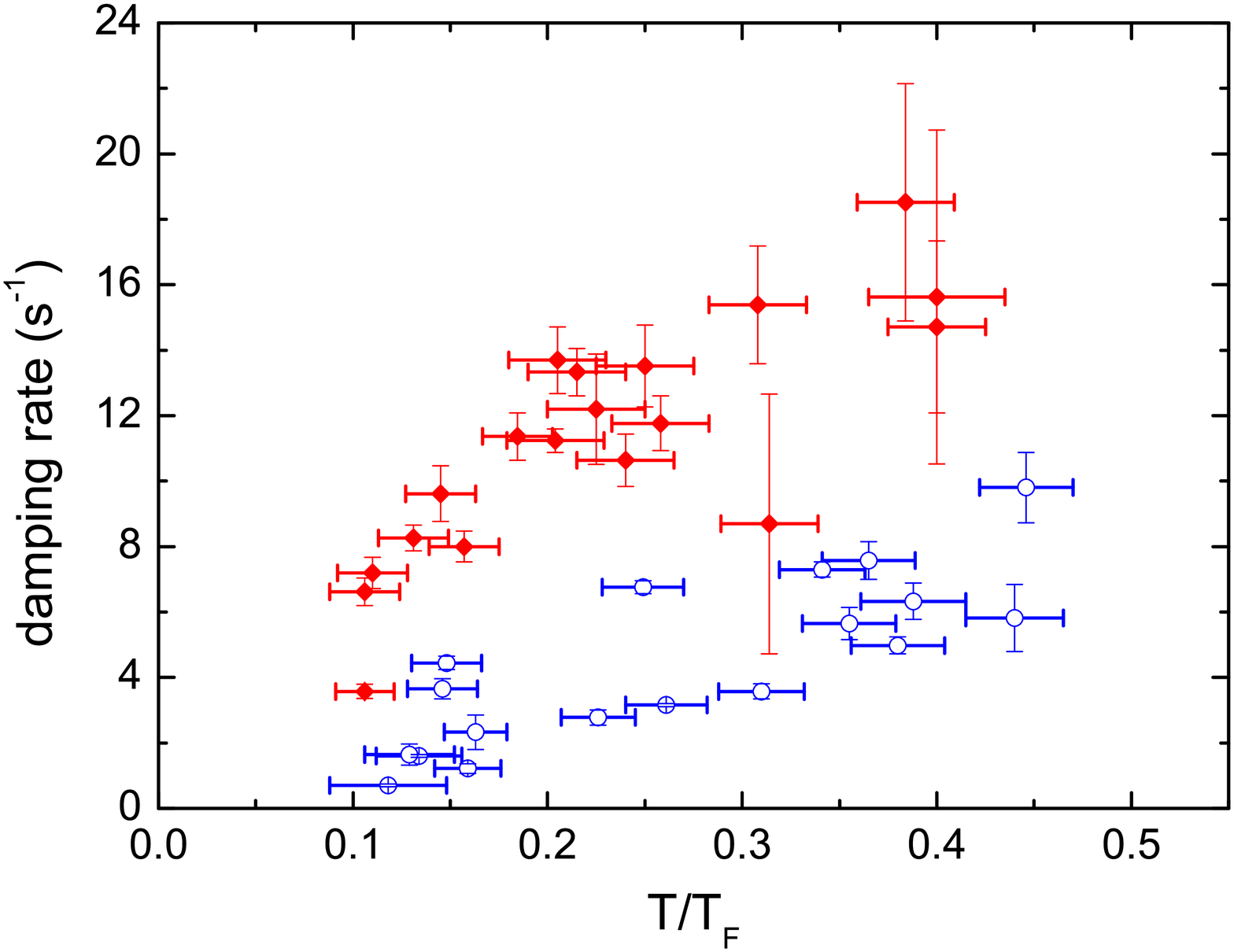}
\caption{\label{fig:damping} (Color online) Measured damping rates of the $k=2$ (blue open circles) and the $k=3$ (red filled diamonds) modes versus temperature.}
\end{figure}

In Fig.~\ref{fig:damping}, we show the damping rates of the $k=2$ (blue open circles) and the $k=3$ (red filled diamonds) modes, measured at various temperatures. The damping rate of the $k=3$ mode is always several times higher than that of $k=2$, and it strongly increases as the cloud gets hotter. It is evident that the situation, where we have observed significant deviations in the mode frequencies ($k=3$ and $T/T_F \gtrsim 0.2$), coincides with the case of highest damping rates (exceeding 10\,s$^{-1}$). It is known that violation of the 1D condition assumed in Sec.~\ref{sec:theory} would result in a damping of the collective oscillation. Therefore, a higher damping rate could naturally be accompanied with a larger deviation from the prediction using the 1D formulation of the Landau's two-fluid model. This supports our interpretation of the observed frequency deviation in terms of dissipative effects.

For an accurate determination of temperature-dependent frequency shifts in our experiment, the $k=2$ mode turns out to be superior to the $k=3$ mode. Our results indicate that the advantage of the latter mode to exhibit larger frequency changes is overcompensated by the larger damping, which introduces larger statistical uncertainties (see error bars in Fig.~\ref{fig:k2k3freq_TTF}) and apparently also systematic errors. However, for larger trap aspect ratios than applied in our experiment, the situation may be different and higher modes may provide further interesting information.

\begin{figure}
\includegraphics[width=0.9\columnwidth]{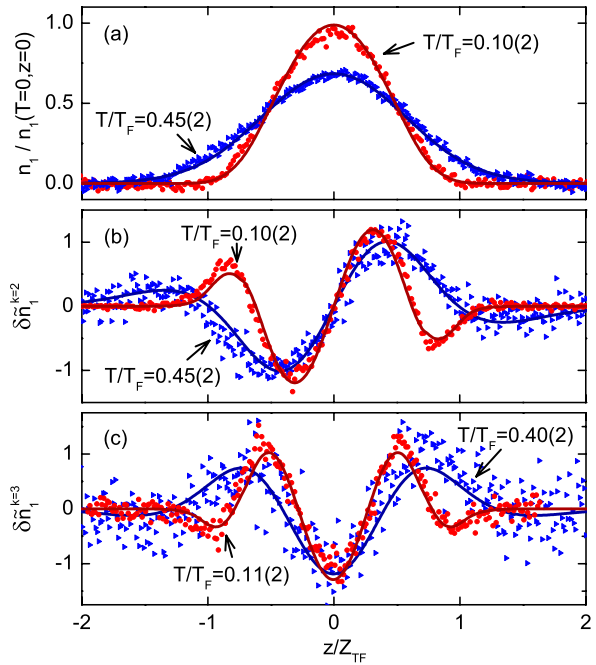}
\caption{\label{fig:profiles} (Color online) Comparison of experimental mode profiles (data points) and theoretical predictions (solid lines) for two different temperatures. In (a), we show the cloud profiles obtained with a 2-ms TOF. The solid lines show the density profiles obtained from the EOS \cite{Ku2012rts} with $T/T_F = 0.10$ and $0.45$. In (b), we show the experimental and theoretical $k=2$ mode profiles at the indicated temperatures. The comparison for the $k=3$ mode is presented in (c). The $z$ scale is normalized to the Thomas-Fermi radius $Z_{\rm TF}$ of the zero-$T$ interacting gas, which is a factor of $\xi^{1/4} \approx 0.78$ \cite{Ku2012rts} smaller than in the non-interacting case.}
\end{figure}

In Fig.~\ref{fig:profiles}(a) and (b), we finally show the observed spatial profiles of the $k=2$ and $k=3$ modes, in comparison to the corresponding theoretical predictions. For both modes we present data sets for the lowest temperature that we could realize ($T/T_F \approx 0.1$) and for the highest temperature explored ($T/T_F \approx 0.45$). In the first case, the situation is deep in the superfluid regime, whereas the second case corresponds to the classical hydrodynamic case. For reference, Fig.~\ref{fig:profiles}(a) shows the corresponding spatial profiles of the unperturbed cloud, from which we obtained the temperature following the method of Sec.~\ref{subsec:EE0} in the Appendix. The agreement between the experimentally observed mode profiles and the theoretical predictions is remarkable. Within the experimental uncertainties and with the mode amplitude being the only fit parameter, we find a perfect match. This again highlights the validity of the theoretical 1D framework and power of our experimental approach to higher-nodal collective modes.

\section{Conclusions and Outlook}
\label{sec:outlook}

We have presented an efficient tool-box to excite and detect higher-nodal axial collective modes in a resonantly interacting Fermi gas. Our results (see also \cite{Tey2013cmi}) reveal the pronounced temperature dependence below and near the superfluid phase transition, which is theoretically predicted in the framework of a 1D two-fluid hydrodynamic model \cite{Hou2013fas}. The observed temperature dependence is a unique feature of higher-nodal modes and has not been observed in any other collective mode studied in Fermi gases so far. The excellent agreement of the experimentally observed mode frequencies with the theoretical predictions provides a stringent test for the validity of this 1D approach and provides an independent confirmation of the recently measured EOS \cite{Ku2012rts} of the resonantly interacting Fermi gas.

We have also reported first studies on the mode damping behavior, which show a strong increase of the measured damping rates with the order of the mode investigated. Dedicated experiments on damping could provide valuable information on the viscosity and the thermal conductivity of the strongly interacting Fermi gas, which may provide further insight into fundamental questions related to viscosity \cite{Schafer2009npf, Cao2010uqv}. A better understanding of damping would also be important to understand the limitations of the theoretical approach \cite{Hou2013fas} applied to describe the modes.

Generalizations of our experiments to first-sound collective modes in the BEC-BCS crossover regime \cite{Giorgini2008tou, Bloch2008mbp, Chin2010fri}) and to spin-polarized Fermi gases \cite{Zwierlein2006fsw, Partridge2006pap, Nascimbene2009coo} will be rather straightforward. A very exciting prospect is the extension to second-sound modes \cite{Taylor2009fas, Hou2013fas}, where the superfluid and the normal component oscillate in opposite phase. A recent experiment \cite{Sidorenkov2013ssa} shows the propagation of second-sound pulses along the trap axis, in agreement with a theoretical description based on the same approach that is used in the present work. This observation may, in principle, be interpreted in terms of a superposition of several second-sound modes, but the selective excitation and observation of individual modes of this kind remains an experimental challenge for future experiments.

\begin{acknowledgments}
We would like to acknowledge the strong theoretical support from Yan-Hua Hou, Lev Pitaevskii, and Sandro Stringari. We thank Mark Ku and Martin Zwierlein for fruitful discussions and Florian Schreck for discussion and experimental support. We acknowledge support from the Austrian Science Fund (FWF) within SFB FoQuS (project No.\ F4004-N16).
\end{acknowledgments}

\appendix*
\section{Temperature determination}\label{appendix}

The recorded {\it in-situ} density profiles $n_1(z)$ are influenced by imperfections in the imaging process. While, under our experimental conditions, the limited resolution and optical aberrations do not play any significant role, we have identified another problem (Sec.~\ref{app:imperfect}) that can considerably affect our thermometry. We here discuss our strategies to circumvent this problem, presenting our two methods (Secs.~\ref{subsec:TTF} and \ref{subsec:EE0}) to extract the temperature from the observed profiles.

\subsection{Imperfections of absorption imaging}
\label{app:imperfect}

\begin{figure}
\includegraphics[width=0.8\columnwidth]{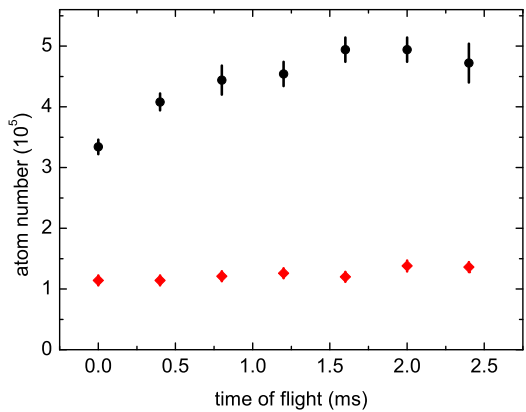}
\caption{\label{fig:Natom}(Color online) Apparent atom numbers obtained with different times of flight while keeping other experimental parameters unchanged. The red diamonds and the black circles represent the data sets with $N = 1.2\times10^5$ and $4.8\times10^5$ atoms, respectively.}
\end{figure}

%When the atoms are tightly packed within the trap or very shortly after release from the trap,
{\it In-situ} absorption images and images taken with a short time of flight (TOF) reveal an apparent reduction of the effective absorption cross section, which predominantly occurs in the denser regions of the cloud. To illustrate this effect, we show in Fig.~\ref{fig:Natom} how the apparent atom number, i.e.\ the atom number obtained under the assumption of the full absorption cross section, depends on the TOF after release from the trap for two experimental settings corresponding to a number of $N = 1.2\times10^5$ atoms (red diamonds) and $4.8\times10^5$ atoms (black circles) in the trap. Only after a TOF of 2\,ms the apparent atom number reaches a constant maximum value, which corresponds to the true atom number. It is evident that our {\it in-situ} imaging underestimates the actual atom numbers by about 15\% for the data set with $N = 1.2\times10^5$, and by about 30\% for $N = 4.8\times10^5$.

We do not fully understand the reason for this reduction, which is clearly related to the high density of the cloud, but it is not related to a `black out' effect, in which the imaging light is completely absorbed by a very dense cloud. In our case, the maximum optical density is about one, which can be properly accounted for by the exponential decrease of the transmission with increasing column density. We speculate that the underlying mechanism is related to multiple scattering of a photon by a few neighboring atoms when the atoms are very closely packed, thereby reducing the effective absorption cross section of the atoms. For our geometry, this effect should be important when the photon's `mean free path' becomes comparable to the radial size of the cloud, which is indeed the case for our experimental conditions. %Alternatively, the effect may be a subtle consequence of aberration effects in our imaging system, occurring when the diameter of the cloud becomes too small

By analyzing profiles obtained for different times of flight we found that the imaging problem mainly affects the center of the cloud, where one finds the largest density. Therefore the problem does not only result in smaller apparent atom numbers, but it also distorts the density distribution $n_1(z)$ of the cloud as obtained by {\it in-situ} absorption imaging. To determine the temperature of the cloud in spite of this problem, we have adopted the two different methods described below. Each method has its own systematic uncertainties. Their accuracies can only be judged \textit{a posteriori} by their mutual agreement and also by their agreement with certain theoretical predictions.

Both methods rely on the accuracy of the total atom number $N$. We obtain this number through a very careful calibration of the imaging process and the imaging system. For imaging, we typically adopt a 10-$\mu$s-long imaging pulse and keep the probe intensity less than 4\% of the saturation intensity. These parameters keep the total number of photon scattering events per atom small, thereby minimizing the photon-recoil-induced Doppler-detuning effect \cite{Hemmer1981ioa, Grimm1988ool} on the light $^6$Li atoms. We estimate the uncertainty in the resulting atom number to be typically about 10\%, in any case below 20\%.

\subsection{\label{subsec:TTF}Wing-fitting method}

\begin{figure}
\includegraphics[width=0.8\columnwidth]{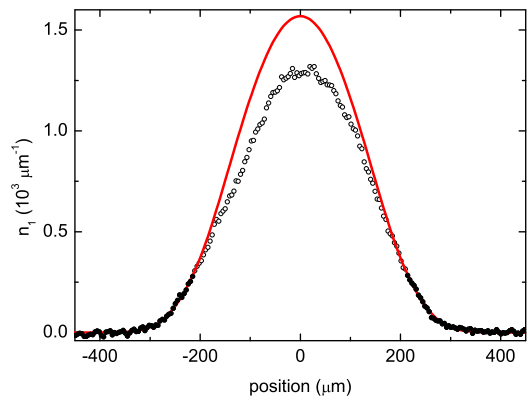}
\caption{\label{fig:wingfit} Typical fit to the wings of the 1D density distribution to determine $T/T_F$ from  Eq.~(\ref{eqn:n1}). The filled circles represent the data used for fitting, and the solid line is the full profile according to the fit. The open circles represent the remaining data that were not used in the fit because of the problem to determine the correct linear density in the center of the cloud. Here a TOF of 600\,$\mu$s was applied, the total atom number is $N = 4.2\times 10^5$, and the temperature is $T/T_F = 0.36$.}
\end{figure}

In the first method, we obtain the cloud's temperature by fitting Eq.~(\ref{eqn:n1}) to the wings of a 1D density profile taken after a 600\,$\mu$s TOF \cite{TOFComment}, with $N$ determined independently from a corresponding 2-ms TOF image. The wings, where $n_1(z)$ is relatively small and the total absorption is weak, are essentially free of the image distortion as described above.
%Consequently, fitting Eqn.~\ref{eqn:n1} to the full profile does not give a correct $T/T_F$.
An example for a typical wing fit is shown in Fig.~\ref{fig:wingfit}. The Figure also illustrates the difference between the true density profile as reconstructed by the wing fit (solid line) and the distorted observed one (data points).

In practice, we adjust the number of data points used for the wing fit until we obtain an atom number very close to the actual value obtained using a sufficiently long TOF. Sometimes, we have to use different numbers of data points in each wing when the profile on the wings is not fully symmetric. This issue is caused mainly by optical aberrations in the imaging system. Nevertheless, we always find the same $T/T_F$ within an uncertainty of 10 to 15\% using this procedure for samples prepared by the same experimental sequence. We also do not observe a clear deviation in the so obtained temperature when varying the TOF from 0 to 600\,$\mu$s. Our experience shows that the accuracy of $T/T_F$ obtained from this method is not very sensitive to slight distortions in the wing profiles. Instead, it depends more crucially on the correct atom number $N$.

\subsection{\label{subsec:EE0}Potential-energy method}

In the second method, we characterize the temperature in a model-independent way that does not require any \textit{a priori} knowledge of the EOS. Based on the virial theorem~\cite{Thomas2005vta} we can obtain the total energy $E$ of the harmonically trapped cloud from its potential energy, which can be calculated from the 1D density profile $n_1(z)$ according to $E=3m\omega_z^2\int_{-\infty}^{\infty} n_1(z) z^2\,dz$.
We define a dimensionless parameter $E/E_0$, where $E$ is normalized to the energy of a noninteracting, zero-temperature Fermi gas with the same number of atoms, $E_0=\frac{3}{4}Nk_BT$.

To obtain an accurate value for $E/E_0$ it is essential to have accurate knowledge of $n_1(z)$. Here we overcome the above-discussed distortion problem by `reconstructing' the correct profile. We measure the density profiles of the cloud for two different times of flight, 600-$\mu$s and 2\,ms. The corresponding profiles $n_1(z,{600\,\mu s})$ and $n_1(z,2\,\mathrm{ms})$ provide complementary information. While the 600-$\mu$s profile provides accurate information on the wings, it underestimates the central part of the cloud. In contrast, the 2-ms profile suffers in the wings from the expansion, but gives accurate information on the central part of the cloud. With an appropriate algorithm to combine this information we can reconstruct the full profile with reasonable accuracy. The reconstructed profile shows approximately the correct atom number and it maintains the wing distribution, which has large weights $\propto z^2$ in the total energy $E$.

\begin{figure}
\includegraphics[width=0.85\columnwidth]{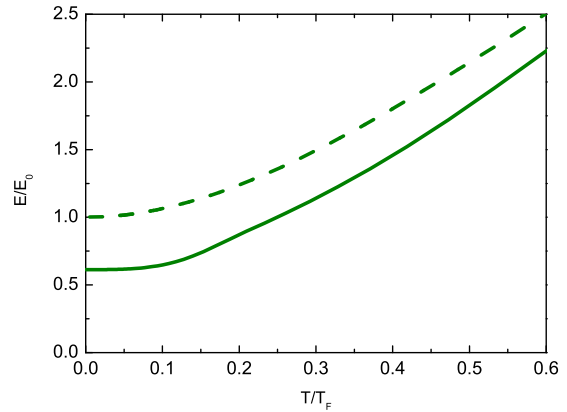}
\caption{\label{fig:EE0VsTTF} Relation between $E/E_0$ and $T/T_F$ for a 3D harmonic trap. The solid and the dashed lines correspond to the results obtained using the EOS from Ref.~\cite{Ku2012rts} and the EOS of the non-interacting Fermi gas, respectively.}
\end{figure}

We can now convert $E/E_0$ into $T/T_F$ using the EOS from Ref.~\cite{Ku2012rts}. Using the universal thermodynamic relations of a resonantly interacting Fermi gas, one can show that $E/E_0$ is related to the $x_0 = \beta\mu_0$ at the center of the trap by
\begin{equation}\label{eqn:epsilon}
\frac{E}{E_0} = \frac{4(2\pi)^{1/2} M_2(x_0)}{3(3\sqrt{2}\pi)^{1/3} M_0^{4/3}(x_0)},
\end{equation}
while the $x_0$ is related to $T/T_F$ by
\begin{equation}\label{eqn:tau}
\frac{T}{T_F}= \frac{(2\pi)^{1/2}}{[24\sqrt{2}\pi M_0(x_0)]^{1/3}}.
\end{equation}
Figure~\ref{fig:EE0VsTTF} shows the conversion between $E/E_0$ and $T/T_F$ for a resonantly interacting Fermi gas in comparison with the ideal non-interacting Fermi gases. For the resonantly interacting Fermi gas at $T=0$, $E/E_0 = \sqrt{\xi}$. Here $\xi$ is the Bertsch parameter, for which Ref.~\cite{Ku2012rts} gives the value $\xi = 0.376(4)$. For the ideal Fermi gas at $T=0$, $E/E_0=1$ by definition.

We finally note that we applied both methods to various data sets to check whether they produce consistent results. In general we find satisfying agreement with each other, as the example of the data set in Fig.~\ref{fig:k2k3freq_TTF}(a) shows. At very low temperatures, the wing-fit method shows a trend to give slightly lower values of $T$ (up to $\sim$10\%) as compared to the potential-energy method. This indicates small systematic uncertainties of our methods.

%\bibliography{../ultracold,highernodallong}

\begin{thebibliography}{46}
\bibitem [{pul()}]{pulseComment}%
  \BibitemOpen
  \href@noop {} {}\bibinfo {note} {The power of the green beam is given by
  $P(t)=P_0\sin(\omega_m t/16)[1-\cos(\omega_m t)]$ for $0<t<16\pi/\omega_m$,
  and zero otherwise. The parameter $\omega_m$ is the modulation frequency and
  $P_0$ represents the peak power.}\BibitemShut {Stop}%
\bibitem [{FFT()}]{FFTComment}%
  \BibitemOpen
  \href@noop {} {}\bibinfo {note} {FFT gives a complex spectrum, $\delta
  n(z,\omega)=\mathrm{Re}[\delta n(z,\omega)] + i\,\mathrm{Im}[\delta
  n(z,\omega)]$. By multiplying a suitable phase factor to $\delta
  n(z,\omega)$, we can transfer all information pertaining to a particular mode
  into one quadrature, obtaining $\delta \tilde{n}(z,\omega)$.}\BibitemShut
  {Stop}%
\end{thebibliography}


%

\end{document}